\documentclass{webofc}
\usepackage{amsmath,amssymb,amstext}
\usepackage{graphicx}
\usepackage{units}
\usepackage{color,xcolor}
\usepackage[absolute,overlay]{textpos}
\usepackage{latexsym,overpic,lipsum}
\usepackage[varg]{txfonts}   
\definecolor{orange}{rgb}{1,.5,.3}
\definecolor{lightyellow}{cmyk}{0,0,0.5,0}
\definecolor{lightred}{rgb}{1,0.5,0.5}
\definecolor{lightgreen}{rgb}{0.8,1.0,0.8}
\definecolor{lightblue}{rgb}{0.5,0.5,1}
\definecolor{darkred}{rgb}{0.8,0,0}
\definecolor{darkgreen}{rgb}{0,0.4,0}
\definecolor{darkcyan}{cmyk}{1,0.3,0.3,0.3}
\definecolor{darkblue}{rgb}{0,0,0.6}
\definecolor{lightbrown}{rgb}{0.7,0.3,0.3}
\definecolor{darkbrown}{rgb}{0.5,0,0}
\definecolor{bluegreen}{rgb}{0,0.5,0.5}%
\newcommand{\bk}{\color{black}}
\newcommand{\rd}{\bk}
\newcommand{\bl}{\bk}

\newcommand{\er}{$\pm$}

\newcommand{\bc}{\begin{center}}
\newcommand{\ec}{\end{center}}
\newcommand{\bi}{\begin{enumerate}}
\newcommand{\ei}{\end{enumerate}}
\newcommand{\tz}{$\to$}

\begin{document}
\title{The Fragmented Glueball: A Personal View}
%
%

\author{\firstname{Eberhard } \lastname{Klempt}\inst{1}\fnsep\thanks{\email{klempt@hiskp.uni-bonn.de}} 
}

\institute{Helmholtz--Institut f\"ur Strahlen-- und Kernphysik der Universit\"at Bonn, Nussallee 14-16, 53115 Bonn, Germany    }

\abstract{%
A coupled-channel analysis has been performed to identify the spectrum of scalar 
mesons. The data include BESIII data on radiative $J/\psi$ decays into $\pi^0\pi^0$,
$K_SK_S$, $\eta\eta$, and $\omega\phi$, 15 Dalitz plots from $\bar pN$ annihilation at 
rest at LEAR, the CERN-Munich multipoles for $\pi\pi$ elastic scattering, the $S$-wave from 
BNL data on $\pi\pi$ scattering into $K_SK_S$, from GAMS data on 
$\pi\pi\to \pi^0\pi^0, \eta\eta$, and $\eta\eta'$,  and NA48/2
data on low-mass $\pi\pi$ interactions from $K^\pm\to\pi\pi e^\pm\nu$ decays. 
The analysis reveals the existence of
ten scalar isoscalar resonances. The resonances can be
grouped into two classes: resonances with a large SU(3) singlet component and those with
a large octet component. The production of isoscalar resonances with a large octet component 
should be suppressed in radiative $J/\psi$ decays. However, in a limited mass range centered at 1900\,MeV, 
these mesons are produced abundantly. Mainly-singlet scalar resonances are produced 
over the full mass range but with larger intensity at 1900\,MeV. The total scalar isoscalar yield 
in radiative decays into scalar mesons shows a clear peak  which is interpreted as the scalar
glueball of lowest mass. }  
\maketitle
\section{Introduction}
\label{intro}
Glueballs, bound states of gluons with any constituent quarks, are a firm prediction of QCD.
Lattice gauge theories predict the lowest-mass glueball to have scalar quantum numbers and
to have a mass of 1710\er50\er80\,MeV~\cite{Chen:2005mg}. Tensor and pseudoscalar 
glueballs are expected well above 2000\,MeV. Analytic approximations to QCD find the scalar glueball 
in the range from 1850 to 1980\,MeV~\cite{Szczepaniak:2003mr,Huber:2020ngt,Rinaldi:2021dxh}.
Several observed scalar mesons have been proposed to contain a large glueball fraction but no 
firm conclusions have been reached so far~\cite{Close:2001ga,Klempt:2007cp,%
Mathieu:2008me,Crede:2008vw,Ochs:2013gi,Amsler:2019inPDG,Llanes-Estrada:2021evz}. 
In a recent paper, a partial-wave analysis of data carrying information on scalar mesons was 
reported~\cite{Sarantsev:2021ein}. Here, I present these results and give my personal interpretation. 

\section{Data and PWA}
\label{sec-1}
\setcounter{figure}{0}\begin{tabular}{cc}
\begin{minipage}[c]{0.7\textwidth}
Decisive for the interpretation of the glueball as part of the spectrum of scalar mesons are
the BESIII data on radiative $J/\psi$ decays. In this process, the primary $c\bar c$ converts into 
one photon and two gluons (see Fig.~1). The gluons interact thus forming a glueball that 
decays into the observed final-state particles.
Figure~\ref{fig-2} shows the invariant mass spectra of $\pi^0\pi^0$, $K_sK_s$, $\eta\eta$,
and $\phi\omega$ in a relative $S$ wave produced in radiative $J/\psi$ decays. In parti-\phantom{zzzzzzzz} 
\end{minipage}
&
\begin{minipage}[c]{0.26\textwidth}
\includegraphics[width=\textwidth]{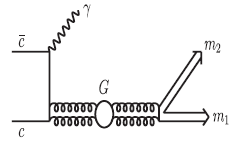}\vspace{-6mm}
\label{fig-1}
\bc
{\bf Figure 1.} Radiative $J/\psi$ decay
\ec
\setcounter{figure}{1}\end{minipage}
\end{tabular}

\clearpage
\begin{figure*}
\centering
\begin{tabular}{cc}
\hspace{-4mm}\raisebox{0.8mm}{\includegraphics[width=0.5\textwidth]{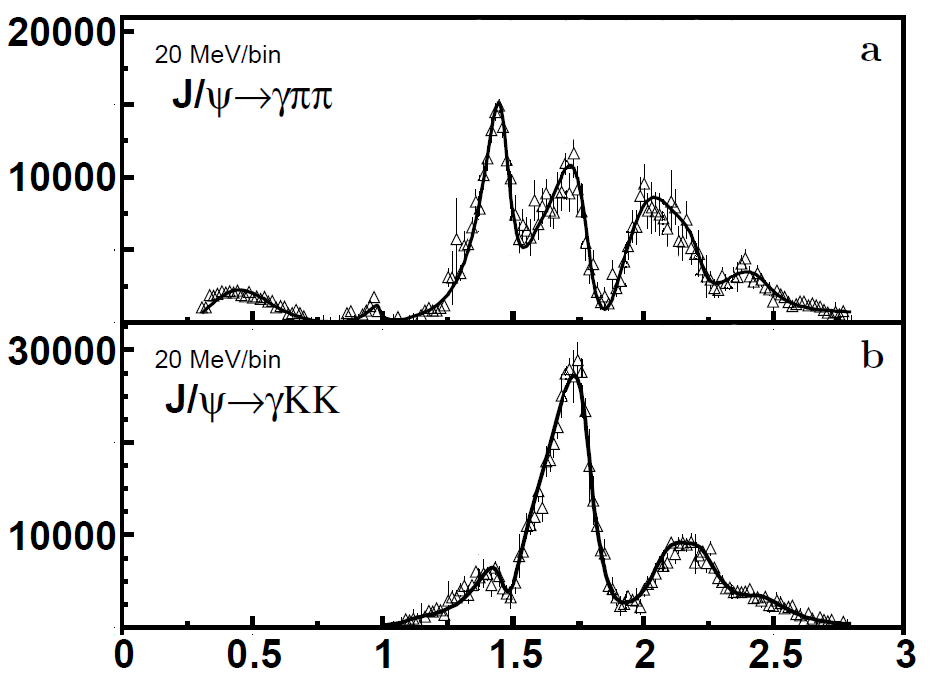}}&
\hspace{-4mm}\includegraphics[width=0.5\textwidth]{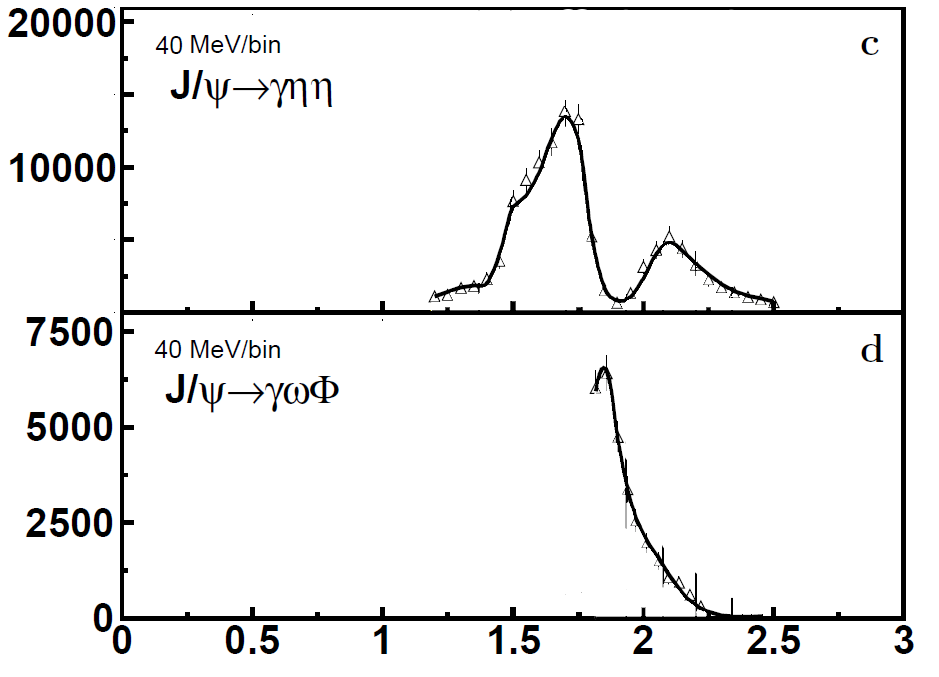}\\
\end{tabular}
\caption{BESIII data on radiative $J/\psi$ data into (a) $\pi^0\pi^0$, (b) $K_sK_s$ (b), $\eta\eta$ (c),
and $\phi\omega$ (d). Shown is the meson-meson $S$-wave contribution.}
\label{fig-2}    \vspace{-5mm}
\end{figure*}
\noindent
cular the $\pi^0\pi^0$ and $K_sK_s$ mass distributions show a series of peaks and valleys. These are often seen at different mass positions emphasizing the importance of
interference effects and the need of coupled-channel analyses.

These data are important for the interpretation. They are, however, not sufficient to constrain the number and the properties of the contributing resonances. Therefore we added several further data: 15
different Dalitz plots from $\bar p N$ annihilation at rest into three pseudoscalar mesons, the
multipoles from the CERN-Munich analysis of elastic $\pi\pi$ scattering data -- these data allow
us to determine the {\it missing intensity} of scalar mesons -- the GAMS data on  $\pi^+\pi^-$ scattering
into $\pi^0\pi^0$, $\eta\eta$, $\eta\eta'$, BNL data on $\pi^+\pi^-$ scattering into $K_sK_s$, and NA48/2
data on low-mass $\pi\pi$ interactions from $K^\pm\to\pi\pi e^\pm\nu$ decays. 

The analysis finds ten scalar isoscalar resonances. Table~\ref{tab-1} shows their masses and widths
and  compares them to PDG values. The five mesons below 1750\,MeV are considered to be established
by the Particle Data Group, $f_0(1770)$ is ``new" (even though evidence had been reported
earlier), the four higher-mass mesons were not regarded as being established. The agreement
between our new values and those reported earlier is remarkable. Table~\ref{tab-2} presents
the yields of scalar mesons in the various final states. The missing intensity is derived from
the CERN-Munich data and compared to the intensities reported for $J/\psi\to\gamma 4\pi$ and
$J/\psi\to\gamma \omega\omega$. The last column gives the sum of all intensity assigned to
a specific resonance. 

\begin{table*}[pb]
\caption{Pole masses and widths (in MeV) of scalar mesons. The RPP values are listed
as small numbers for comparison.}
\label{tab-1}       
\renewcommand{\arraystretch}{1.3}
\centering
\begin{tabular}{cccccc}
\hline\hline
Name       & \rd$f_0(500)$     &\rd$f_0(1370)$     &\rd$f_0(1710)$  &\rd$f_0(2020)$                 &\rd$f_0(2200)$     \\\hline
$M$          &  410\er 20    & 1370\er 40      &1700\er 18    & 1925\er 25                  & 2200\er 25  \\[-1.5ex]
              &\scriptsize 400\tz 550 &\scriptsize 1200\tz 1500&\scriptsize 1704\er12&\scriptsize 1992\er 16              &\scriptsize 2187\er 14\\
$\Gamma$   & 480\er30       &  390\er 40       & 255\er 25    & 320\er 35                   & 150\er 30\\[-1.5ex]
               &\scriptsize 400\tz 700 & \scriptsize 100\tz 500  &\scriptsize 123\er 18 & \scriptsize 442\er60               &\scriptsize $\sim 200$ \\
\hline\hline
Name       &\bl $f_0(980)$     &\bl$f_0(1500)$     &\bl$f_0(1770)$  &\bl$f_0(2100)$                 &\bl$f_0(2330)$     \\\hline
$M$         &1014\er 8      &  1483\,\er\,15    &1765\er 15   &2075\er 20                   &2340\er 20\\[-1.5ex]
               &\scriptsize 990\er 20   & \scriptsize 1506\,\er\,6&                  &\scriptsize 2086$^{+20}_{-24}$&\scriptsize$\sim$2330\\
$\Gamma$&71\er10          & 116\er 12        & 180\er 20    & 260\er 25                   & 165\er 25\\[-1.5ex]
               &\scriptsize 10\tz 100   & \scriptsize 112\er 9     &                  & \scriptsize 284$^{+60}_{-32}$ &\scriptsize 250\er 20\\
\hline\hline\vspace{-2mm}
\end{tabular}
\end{table*}
\clearpage
The sum of the yields of $f_0(1710)$ and $f_0(1770)$ should be compared to the RPP entry
for $f_0(1750)$. RPP reports yields for $f_0(2100)$ and $f_0(2200)$. The yields can be compared
to the sum of three resonances found by us. Above 1900\,MeV, no data on $\pi\pi$ elastic scattering are
known. The total $\gamma 4\pi$ intensity is therefore distributed between the three states
observed in the 2000 to 2200\,MeV range.

\begin{table*}
\caption{Yields  of scalar mesons in radiative $J/\psi$ decays (in units of $10^{-5}$). 
The RPP values are listed
as small numbers for comparison.}
\label{tab-2}       
\renewcommand{\arraystretch}{1.4}
\centering\small
\begin{tabular}{ccccccccc}
\hline\hline\hspace{-4mm} $BR_{J/\psi\to\gamma f_0\to}$ & $\gamma{\pi\pi}$&$\gamma{K\bar K}$&$\gamma{\eta\eta}$&$\gamma{\eta\eta'}$&$\gamma{\omega\phi}$&\multicolumn{2}{c}{missing}&total\\[-1ex]
                        &&&&&&\footnotesize$\gamma{4\pi}$&\footnotesize$\gamma\omega\omega$\\\hline\hline\\[-4ex]
$ f_0(500)$& 105\er20   &5\er 5&4\er 3 &$\sim$0&$\sim$0&\multicolumn{2}{c}{$\sim$0}&114\er21\\\hline
$f_0(980)$&1.3\er 0.2  &0.8\er0.3&$\sim$0&$\sim$0&$\sim$0&\multicolumn{2}{c}{$\sim$0}&2.1\er0.4\\\hline
$f_0(1370)$&38\er 10  &13\er4&3.5\er1&0.9\er 0.3&$\sim$0&\multicolumn{2}{c}{14\er5}&69\er12\\[-1ex]
                        &&\scriptsize 42\er15&&&&\scriptsize 27\er9\\\hline
$ f_0(1500)$&9.0\er1.7 &3\er1&1.1\er0.4&1.2\er0.5&$\sim$0&\multicolumn{2}{c}{33\er8}&47\er9\\[-1ex]
                        &\scriptsize 10.9\er2.4&\scriptsize$2.9$\er1.2&\scriptsize $1.7^{+0.6}_{-1.4}$&\scriptsize 6.4$^{+1.0}_{-2.2}$&&\scriptsize 36\er9\\\hline
$f_0(1710)$&6\er2 &23\er8&12\er4&6.5\er2.5&1\er 1&\multicolumn{2}{c}{7\er3}&56\er10\\
$f_0(1770)$&24\er8 &60\er20&7\er 1&2.5\er1.1&22\er4&\multicolumn{2}{c}{65\er15}&181\er26\\[-1ex]
\scriptsize $f_0(1750)$&\scriptsize 38\er 5    &\scriptsize $99^{+10}_{\ -6}$&\scriptsize $24^{+12}_{\ -7}$&&\scriptsize 25\er 6&\scriptsize 97\er18&\scriptsize 31\er10\\\hline
$f_0(2020)$&42\er 10 &55\er25&10\er10&&&\multicolumn{2}{c}{\scriptsize (38\er13)}&145\er32\\
$f_0(2100)$&20\er 8 &32\er20&18\er15&&&\multicolumn{2}{c}{\scriptsize  (38\er13)}&108\er25\\
$f_0(2200)$&5\er2 &5\er5&0.7\er0.4&&&\multicolumn{2}{c}{\scriptsize  (38\er13)}&49\er17\\[-1ex]
\hspace{-4mm}\scriptsize$ f_0(2100)/f_0(2200)$&\scriptsize 62\er10 &\scriptsize 109$^{+\ 8}_{-19}$&\scriptsize 11.0$^{+6.5}_{-3.0}$&&&\scriptsize 115\er41\\\hline
$f_0(2330)$&4\er2 &2.5\er0.5&1.5\er0.4&&&&&8\er3\\[-1ex]
&&\scriptsize 20\er3&&&&\\
\hline\hline
\end{tabular}
\end{table*}
\section{Interpretation}
\hspace{-2mm}\begin{tabular}{cc}
\begin{minipage}[c]{0.52\textwidth}
The mesons in Table~\ref{tab-1} are grouped pairwise. There are two mesons close in
mass followed by a mass gap. Figure~\ref{fig-3} shows  the squared mass values 
of the ``higher-mass" and the ``lower-mass" mesons as a function of a consequtive
number. The squared masses depend about linearly on the consecutive number.  The numbering 
starts with (-1) since we do not want to imply that $f_0(500)$ is the ground state of the
scalar-isoscalar-meson spectrum. The ``lower" and ``higher"-mass mesons are separated 
by the mass-square difference between~$\eta$~and~$\eta'$~mesons  
but the spectrum is inverted. This is expected \phantom{xxxxx}
\\[-4.5ex]
\end{minipage}
&
\begin{minipage}[c]{0.46\textwidth}
\begin{overpic}[width=\textwidth]{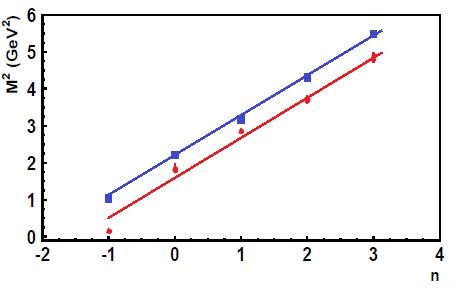}
\put(15,47){\boldmath $\delta M^2 = M_{\eta'} ^2 - M_\eta ^2$}
\put(63,47){\boldmath $\downarrow$}
\put(63,37){\boldmath $\uparrow$}
\put(50,25){\boldmath Slope: 1.08/GeV$^2$}
\end{overpic}\\
{\bf Figure 3.} Squared masses of mainly-octet and 
mainly-singlet scalar isoscalar mesons as functions of a consecutive
number.\vspace{5mm}
\label{fig-3}       
\end{minipage}
\end{tabular}
for instanton-induced interactions~\cite{Klempt:1995ku}. 
For scalar mesons,
$f_0(500)$ is mainly singlet, $f_0(980)$ is mainly octet~\cite{Oller:2003vf,Klempt:2021nuf}.
The decay pattern identifies  $f_0(1370)$ as mainly singlet, $f_0(1500)$ as mainly~octet~state.~Thus we conjecture that the $q\bar q$ component of 
the ``higher-mass" mesons is mainly~in~the~octet,~the  ``lower-mass" mesons in the singlet configuration.

\hspace{-7.2mm}\begin{tabular}{cc}
\begin{minipage}[c]{0.42\textwidth}
Mainly-octet mesons should not be produced in radiative $J/\psi$ decays but they are. 
Figure~4 shows the total yield of scalar isoscalar mesons in radiative $J/\psi$ decays
(from Table~\ref{tab-2}) as a function of the invariant mass. Open circles represent the yield
of ``higher-mass", full squares of ``lower-mass" scalar mesons. Even though the quark content 
of the ``higher-mass" scalar mesons is supposed to be mainly in the octet configuration, their
is significant production in the mass range from 1500 to 2300\,MeV. This high yield requires a
significant singlet contribution
\end{minipage}
&
\begin{minipage}[c]{0.54\textwidth}
\includegraphics[width=\textwidth]{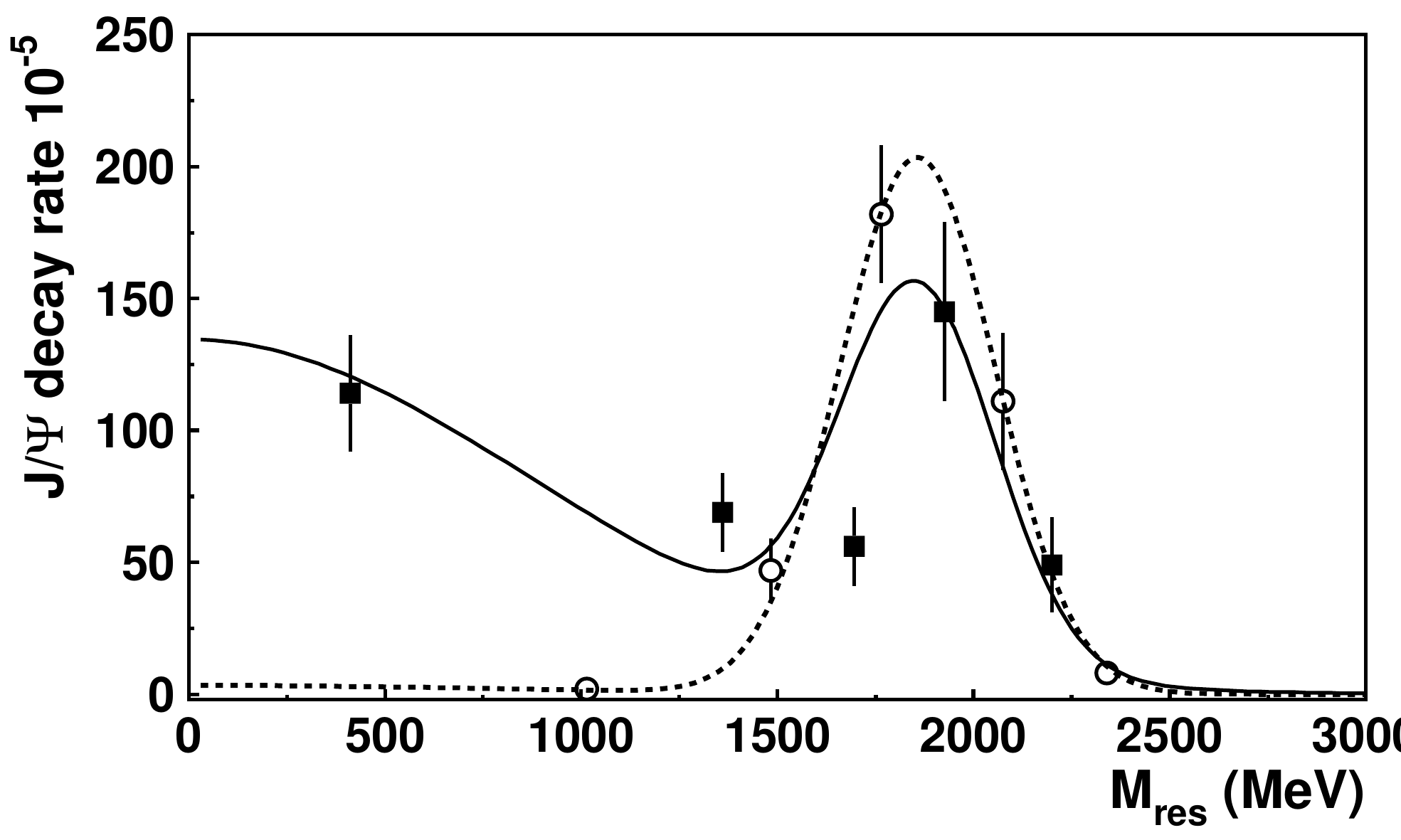}\\
{\bf Figure 4.} Squared masses of mainly-octet and 
mainly-singlet scalar isoscalar mesons as functions of a consecutive
number.\\
\label{fig-4}       
\end{minipage}
\vspace{1mm}\end{tabular}
in their wave functions. No selection rule
forbids SU(3) singlet scalar mesons to be produced in radiative $J/\psi$ decays, they are produced 
in the full mass range but their production 
is enhanced in the same mass range. We interpret this enhancement as scalar glueball.\\[-1ex]

\hspace{-8mm}\begin{tabular}{cc}
\begin{minipage}[c]{0.55\textwidth}
\begin{overpic}[width=0.99\textwidth]{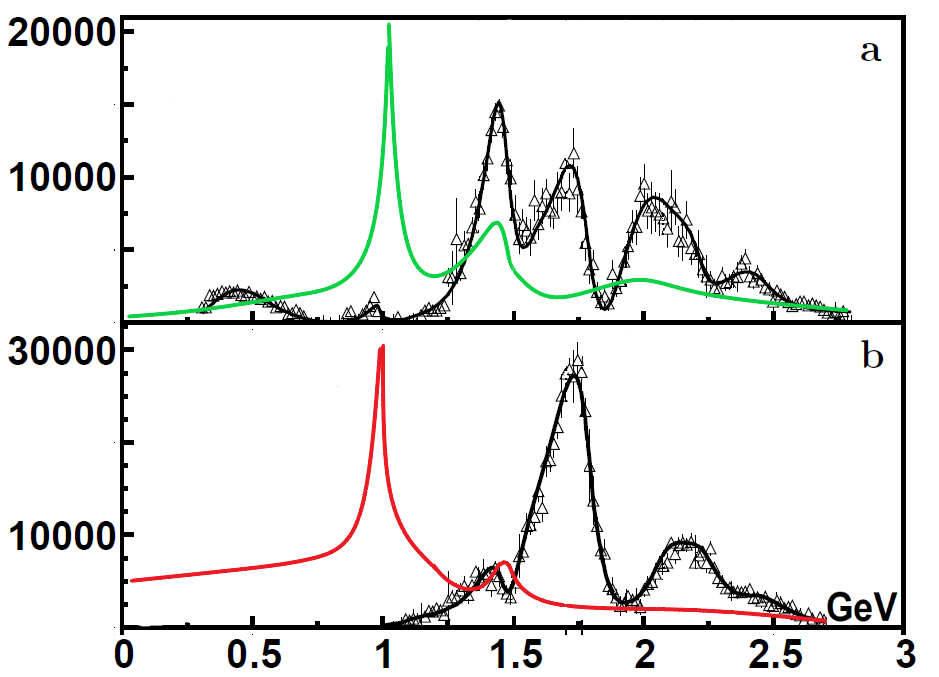}
\put(15,64){\boldmath Pion form}
\put(15,59){\boldmath factor}
\put(15,32){\boldmath Kaon form} 
\put(15,27){\boldmath factor}
\end{overpic}\\
\includegraphics[width=0.95\textwidth]{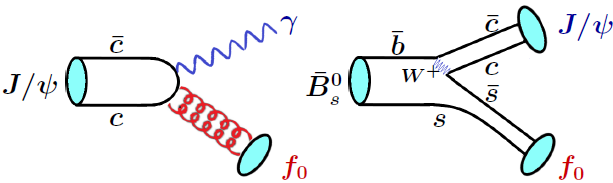}
{\bf Figure 5.} Comparison of scalar-meson production in radiative $J/\psi$ decay and in
$\bar B^0 _s\to J/\psi f_0$. \\
\end{minipage}
&
\begin{minipage}[c]{0.42\textwidth}
\label{fig-4}
This interpretation is confirmed when data on $J/\psi\to \gamma f_0$ and the LHCb data
$\bar B^0 _s\to J/\psi f_0$~\cite{Aaij:2014emv,Aaij:2017zgz} are compared. 
In  $J/\psi$ decays, 
two gluons convert into the scalar meson
that decays into the final-state particles. A montainous landscape is generated with
high peaks and deep valleys. For $\bar B^0 _s$ decays, the form factors from~\cite{Ropertz:2018stk}
are shown in Fig.~5. Their square
is proportional to the cross section. Very little intensity is observed. This is particulary striking
when two kaons are seen in the final state (the kaon form factor). In $\bar B^0 _s$ decays
an $s\bar s$ pair is generated in the initial state but no resonance is produced even though
there is a strong $K\bar K$ peak when two gluons are in the initial state. The two montains
in the $K\bar K$ invariant mass at 1700 and 2200\,MeV are not generated -- or at
most weakly -- by  
\end{minipage}
\vspace{1mm}\end{tabular}
$s\bar s$ pairs but abundantly in gluon-gluon interactions. 

Radiative $J/\psi$ decays produce scalar isoscalar mesons abundantly. Those mesons that
are supposed to have a mostly-octet $q\bar q$ structure (and that should not be produced 
in radiative $J/\psi$ decays) are produced but only in a limited 
mass range, with a clear peak at about 1900\,MeV. Mostly-singlet isoscalar mesons are 
produced in the full mass range but the yield is enhanced at about 1900\,MeV. This peak
is not seen in $\bar B^0 _s$ decays, it is seen only with two gluons in the initial state. 
The enhancement can be fit with a Breit-Wigner resonance.
The fit returns 
\bc
  $\rm M_{glueball}=(1865\pm 25)\,MeV$, $\Gamma_{\rm glueball} =\rm  (370\pm 50^{+30}_{-20})\,MeV$\\[0.5ex]
$\rm Y_{J/\psi\to \gamma G_0}=(5.8\pm1.0)\cdot 10^{-3}$
\ec
This is the largest yield in radiative $J/\psi$ decays into a meson. It is the expected scalar glueball.

\section{Discussion}
Is this really the expected scalar glueball? At least one aspect is completely unexpected. There are ten
scalar isoscalar resonances all falling on a linear $(n, M^2)$ trajectory. All find a spectroscopic 
identification within the quark model -- except perhaps $f_0(500)$ and $f_0(980)$. There is no
extra state, no intruder that enters the spectrum, mixes with other states and increases the
number of observed states by one. In Fig.~\ref{fig-3}, there is no additional resonance.
How can that be?  The scalar glueball seems to make up a fraction of the wave function of
scalar isoscalar mesons without increasing the number of states. 

We can write down the quark decomposition of the wave function of a mainly-octet scalar meson:
\begin{align}
f_0 &= {\bl\alpha_1\frac{1}{\sqrt 6}(u\bar u + d\bar d-2s\bar s)} +{\bl \alpha_2\frac{1}{\sqrt 6}(u\bar u s\bar s + d\bar ds\bar s- 2u\bar u d\bar d)}\nonumber\\
              &+ \alpha_3\cdot\rm (meson-meson\ cloud) +\rd \alpha_4 (gg) +\rd \alpha_5(q\bar qg)\nonumber\\
&+  \rm\bk and\ some\ singlet\ contribution:\ \ {\bk\{\alpha'\frac{1}{\sqrt 3}(u\bar u + d\bar d+s\bar s) +\beta'\frac{1}{\sqrt 3}(u\bar u s\bar s + 
d\bar ds\bar s +u\bar u d\bar d)\}} \nonumber
\end{align}
These are five different octet contributions, five Fock components. These could all be realized
independently. They could mix, but the number of ground states -- and the number of states
in every excitation level -- could be five. This seems not to be the case. The number of states
is rather one per excitation level and not five. The large number
of different Fock components does obviously not lead to a large number of states but only to components
in the wave functions. 

\section{Summary}
The BESIII collaboration reported data on radiative $J/\psi$ decays with unprecedented statistics.
The data reveal high intensities in the yield of scalar mesons. 
 The data can be fit with ten scalar isoscalar resonances.
 The scalar resonances can be grouped into a class of mainly-singlet
and mainly-octet states.
The two groups fall onto linear $(n, M^2)$-trajectories.
 Octet scalar isoscalar resonances are produced mainly in the 1700 - 2100\,MeV  mass range.
 Singlet scalar resonances are produced over the full mass range but their intensity peaks
in the 1700 - 2100\,MeV  mass range.  
 The enhanced production of scalar mesons in the 1700 - 2100\,MeV  mass range is due
to gluon-gluon in the initial state.
The peak is the scalar glueball of lowest mass. The glueball does not intrude the spectrum of scalar
states as additional resonance. It contributes to the wave function of several scalar isoscalar
resonances.

\end{document}